\documentstyle[aps,twocolumn,psfig]{revtex}

\begin{document}

%\draft command makes pacs numbers print
%\draft

\title{Comparison of 32-site exact diagonalization results
and \\ ARPES spectral functions for the AFM insulator ${\rm Sr_2CuO_2Cl_2}$}


\author{P. W. Leung\cite{email}}
\address{Dept. of Physics, Hong Kong University of Science and Technology,\\
Clear Water Bay, Hong Kong}
\author{B. O. Wells\cite{present}}
\address{Dept. of Physics and the Center for Materials Science
and Engineering,\\ Massachusetts Institute of Technology, Cambridge,
Massachusetts, U.S.A.}
\author{R. J. Gooding}
\address{Dept. of Physics, Queen's University,
Kingston, Ontario, Canada.}

\date{\today}
\maketitle
%\newpage
\begin{abstract}

We explore the success of various versions of the one--band $t-J$ model 
in explaining the full
spectral functions found in angle--resolved photoemission spectra 
for the prototypical, quasi two--dimensional, tetragonal, antiferromagnetic 
insulator ${\rm Sr_2CuO_2Cl_2}$. After presenting arguments justifying
our extraction of $A({\bf k},\omega)$ from the experimental data,
we rely on exact--diagonalization results from studies of a square 32--site 
lattice, the largest cluster for which such information is presently available,
to perform this comparison.
Our work leads us to believe that (i) a one--band model that includes
hopping out to third-nearest neighbours, as well three--site, 
spin--dependent hopping, can indeed explain not only the dispersion relation, 
but also the quasiparticle lifetimes --- only in the neighbourhood of 
${\bf k} = (\frac{\pi}{2},0)$ do we find disagreement; (ii)
an energy--dependent broadening function, 
$\Gamma (E) = \Gamma_0 + A~E$, is important in accounting
for the incoherent contributions to the spectral functions.

\end{abstract}
%\newpage

\section{Introduction}

The study of strongly correlated electronic systems in two dimensions (2D)
has been very active, particularly since the discovery of the high $T_c$ 
superconductors.
However, while there have been great improvements in the reliability 
and scope of experimental measurements, this has not led to
agreement as to the correct description of either high temperature
superconductivity, in particular, or two--dimensional correlated systems, in general. 
Quite the opposite is in fact apparent 
--- certain experimental results are consistent with one phenomenology,
but not all experiments seem to support the same physical concepts,
and as of yet no one theory is capable of explaining all the data.
One avenue that may alleviate this confusion and allow for a consensus
to develop follows from the study of {\em ideal experimental systems}. 
By this we imply certain compounds are amenable, for various physical as well
as circumstantial reasons, to a complete characterization. Further, 
these compounds are such that a relatively simple theoretical model,
believed to be appropriate for that material, can 
be defined, and with suitable ingenuity we can expect to carry out
a fully quantitative comparison of this model and the experimental data.

A potential paradigm of 2D antiferromagnetic (AFM) insulators, a category
which includes the parent compounds of high T$_c$ superconductors,
is the ${\rm Sr_2CuO_2Cl_2}$ compound \cite {johnston_review}.
The features that make this compound ideal are: (i) it remains
tetragonal down to 10 K; (ii) it has an extremely weak coupling between
${\rm CuO_2}$ planes, and is thus a more exaggerated quasi--2D system
than most high $T_c$ systems; and (iii) this material seems to be 
extremely difficult to dope away from half filling using normal
preparation techniques (a summary of this behaviour may be found in
Ref. \cite {johnston_review}). Further, at present our 
experimental understanding of this material is nearly complete.
Its structure has been determined by x--ray and neutron
diffraction\cite{miller90,vaknin90} to be body--centered--tetragonal
(${\rm K_2NiF_4}$ type), and no transition to the orthorhombic phase
is observed down to at least 10 K.
This compound has a three--dimensional AFM structure and a 
N\'eel temperature $T_N=251\pm5$ K. In fact, this compound is considered to 
be the best available experimental realization of the $S = \frac{1}{2}$
2D square lattice Heisenberg antiferromagnet (2DSLHA) \cite{vaknin90,greven94},
with nearest neighbour exchange $J=125\pm6$ meV.
Lastly, resistivity measurements show that it is strongly 
insulating \cite{miller90}.

The (essentially) undopable character of this compound makes it an ideal 
candidate for angle--resolved photoemission (ARPES) experiments. Such
studies allow for the measurement of the
spectral function $A^{(-)}({\bf k}, \omega)$ (simply abbreviated as 
$A({\bf k},\omega)$ from now on) of a {\em single hole} (left behind by 
the creation of a photoelectron) propagating in a ${\rm CuO_2}$ plane, 
and such data has recently become available \cite{wells}.
The results from this experiment thus provide an ideal testing ground for 
theories that purport to mimic the low--energy physics of ${\rm CuO_2}$ 
planes; {\em e.g.}, the most 
common model, the so--called $t-J$ model, describes spinless vacancies 
hopping
between neighbouring sites in a $S = \frac{1}{2}$ 2DSLHA. If this model is 
indeed
a good representation of a single carrier moving in this plane, it should 
reproduce 
the {\em full spectral functions} measured in Ref. \cite{wells}. The point
of this paper is to demonstrate how well such models of carrier
motion in strongly correlated systems represent the spectral properties
via a comparison with the full, measured spectral functions of 
Ref. \cite{wells}.

%\newpage
\subsection{Theoretical Comparisons}

Not surprisingly, the ARPES results of Ref. \cite {wells} have been the subject
of intense theoretical studies. For example, it has been noted that the 
single--hole $t-J$ model cannot reproduce the experimental band structure 
everywhere
in the Brillouin zone, particularly along the $(0,0)$ to $(\pi,0)$,
and the $(\pi,0)$ to $(0,\pi)$ directions \cite{wells}. Instead,
at the very least hoppings beyond near--neighbour must be added
to the $t-J$ model \cite {ehasym,nazarenko,kyung,sasha,wheatley}. 
The recognition of the importance of further hoppings has also been obtained
in a spin--density wave treatment of the one--band Hubbard 
model \cite {chubukov}.
Work on the more physically plausible three--band  model has shown 
that this model is superior in reproducing the experimental band structure in 
all regions of the first Brillouin zone \cite{nazarenko,starykh,vos}. 
However, 3--band models may be unnecessarily complicated, and thus efforts
to find a useful and quantitatively accurate one--band model are warranted.

The above comparisons were based on theoretically determined and experimentally
inferred dispersion relations. Of course, experimentally much more information
is available, namely, the entire spectral function is known at many wave 
vectors. 
Further, it would be best if the comparison of theory and experiment was based 
on exact theoretical results. This is not possible unless one treats the 
strong coupling
Hamiltonians numerically. Recently we have completed the first exact 
diagonalization 
study on a 32--site square lattice for the $t-J$ model \cite{lg95}, 
a formidable
task (computationally speaking). This achievement is significant because exact
diagonalization results obtained with the smaller $4 \times 4$ fully square 
cluster are subject to the oddity that this lattice is equivalent to the 
four--dimensional hypercube (in the absence of interactions longer than 
nearest neighbour). 
Other exact results have been obtained on smaller, non--square clusters, or on
clusters that lack important wave vectors (namely, those along the AFM Brillouin
zone boundary); a summary of many of these results is contained in a recent
review \cite {elbioRMP}. And lastly and perhaps most importantly, while 
analytical work based on the self--consistent Born approximation for the 
$t-J$ model has been shown to be very reliable, at least as far as the 
dispersion relations are concerned \cite {lg95}, so far no analogous 
demonstration 
of the accuracy of this approach to include further--than--nearest--neighbour 
hoppings of the $t-J$ model has been published. Thus, we consider our 
numerical, 
unbiased, 32--site exact diagonalization approach to a precise manner in 
which these models can be compared to experiment.

In the present paper we continue our numerical studies on
the 32--site square lattice, and present a comparison of the full spectral 
functions for the so--called 
$t-t^{\prime}-t^{\prime\prime}-J$ model, 
a generalization of the simpler
$t-J$ model that includes further hopping processes, as well as three--site,
spin--dependent hopping. Various groups have claimed that this model accurately
fits the dispersion relations found in the AFM insulator ${\rm Sr_2CuO_2Cl_2}$,
and such claims are something that we can critique with our numerical results.
Here, besides continuing these comparisons with our numerical work,
we propose an energy dependence of the self--energy for the propagating
spin polarons.

%\newpage
\subsection {Experimental Analysis}

Angle--resolved photoelectron spectroscopy (ARPES) has proven to be an 
invaluable experimental tool for the understanding of the electronic 
structure of the cuprate superconductors and related correlated--electron
materials. The analysis of the ARPES results depends on the emission 
process being primarily direct transitions so that the spectra 
reflect the spectral function $A({\bf k},\omega )$. This has
allowed for the extensive use of ARPES to determine energy versus {\bf k} 
relations for the highest energy band in the cuprates, and particularly 
for determining the Fermi surfaces of the metals. Most comparisons between 
calculations and experiment have been performed by extracting a quasiparticle 
energy at each {\bf k} from both the experimental spectra and the calculated 
$ A({\bf k},\omega) $ and comparing the dispersions thus derived. However, since 
both experiment and theoretical calculations represent the spectral function, 
it should be more instructive to compare the two directly rather than 
just comparing the derived dispersions. A direct comparison of spectral
functions has its own difficulties, as are outlined below, but hopefully will 
allow a more complete evaluation of the spectral function derived from the Green's 
function and thus avoid the possible errors involved in determining the
experimental quasiparticle dispersion.  

For two--dimensional materials such as the cuprates, the inherent 
lineshape of the ARPES spectrum may reveal 
information about the nature of the elementary excitations. 
Many previous ARPES studies have examined the lineshape of the spectral 
functions. (For a review of this subject, see Ref. \cite {shen_review}; 
for particularly 
relevant experimental work, see \cite {previous_expts}.) 
In these studies the assumption
is made that direct transitions dominate the spectral intensity, and 
thus one may model the spectral response with a single peak, and it 
has been proposed that the nature of this peak should be characteristic 
of a Fermi liquid, a marginal Fermi liquid, or 
some other microscopic model. Since these fits have all been to a 
single peak they represent only the coherent part of the spectrum.
Most numerical calculations of $ A({\bf k},\omega) $ for the cuprates indicate 
that there is a great deal of spectral weight in the incoherent parts of 
the excitations. Therefore, attempting to fit the cuprate ARPES 
spectra with a single peak of any shape is inherently flawed. 
In this work we make no attempt to deduce a specific lineshape, 
or broadening function, but merely use a few simple fitting procedures
as discussed below. The essential point is that for the purposes of fitting 
the experimental data, a correct form for $ A({\bf k},\omega) $ 
is more important than the detailed lineshape of a single peak.

        The planar copper oxides are well suited for analysis
since in most of these materials there is a single highest energy 
occupied state (band) which is well separated in energy from any 
other bands. Experimentally, ${\rm Sr_2CuO_2Cl_2}$ has some particular 
advantages for the present study in addition to those mentioned above. 
It is easy to prepare good surfaces
for photoemission since the material cleaves in a manner 
similar to mica. In addition, the fact that the oxygen content is 
fixed to the stoichiometric value gives us confidence that the 
actual surface under study truly represents a CuO$_2$ plane 
with a single hole per copper site. Thus, while the 
lineshape and broadening appropriate for a particular peak in 
$A({\bf k},\omega)$ may be more difficult to determine than 
for the conducting cuprates, it should be more straight forward to 
determine the correct Greens function for comparison.

Our paper is organized as follows. Firstly, in \S~\ref{expt} we discuss
the extraction of the lineshapes of the spin--polaron band from
the ARPES data.
Then, in \S~\ref{theory} we discuss the various 
modifications to the $t-J$ model necessary to improve the success of
this model in explaining the ARPES results. In \S~\ref{comparison}
we present our numerical results and the comparison of our work to the 
experimental spectral functions; conclusions are provided in \S~\ref{concl}.

%\newpage
\section{Extraction of $A({\lowercase{\bf k}},\omega)$ from ARPES data}
\label{expt}

The collection of the relevant data has already been discussed in Ref.
\cite {wells}, and in that paper the manner in which 
the quasiparticle dispersion relation was obtained from the ARPES data
was presented. Here we focus on the quasiparticle lineshapes.

We propose that the experimental data corresponds to such a 
direct--transition--dominated $A({\bf k},\omega)$,
plus a background which influences the low energy part
of the data and gives rise to an artificially anisotropic
peak shape.  We need to subtract off this background from the
experimental data before they can be compared to the theoretical
$A({\bf k},\omega)$.  We have little theoretical guidance for the 
background function, but we find that a single Gaussian peak 
plus a constant fits the low energy tail of the experimental data very
well. In principle, the background can also include a function
linear in $\omega$ to account for inelastic scattering, and
such a linear function should be {\bf k}--independent.  However,
the best fit at every {\bf k} gives different linear functions.
Since this term is quite small in all cases, we have
chosen to simply set it to be zero.
Since this background function describes
the low energy valence band, we allow it to be {\bf k}--dependent.
We determine this background function at each {\bf k} by 
least--square fitting, and subtract it off from the experimental data.  

The corrected spectra have the following characteristics:
(i) the peaks broaden and weaken as {\bf k} moves away from 
the valence band maximum (VBM) at $(\frac{\pi}{2},\frac{\pi}{2})$,
(ii) the peaks at their sharpest are broad and asymmetrical,
and (iii) although the quasiparticle dispersion is flat along the $(0,0)$
to $(\pi,0)$ direction, the spectral weight is a maximum at 
$(\frac{\pi}{2},0)$.

\begin{figure}
\centerline{\psfig{figure=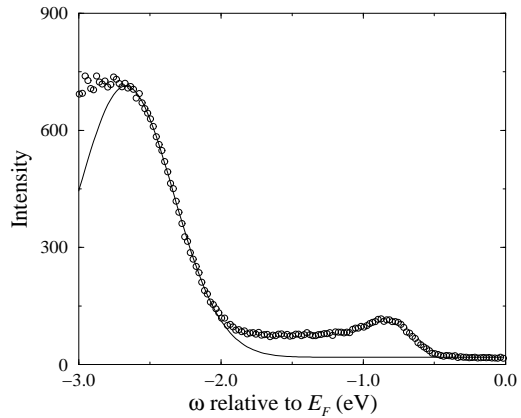,width=3.in}}
\caption{The background function for the main valence band of the insulator 
(solid line) superimposed on the ARPES data, represented by open circles 
($\circ$), for 
${\bf k} = (\frac{\pi}{2},\frac{\pi}{2})$. The remaining spectral weight
is associated with the intrinsic spectral response, coherent as well
as incoherent, of a single hole in a 2D AFM.}
\label{extraction}
\end{figure}

As an example of the application of this procedure, in Fig.~\ref{extraction}
we show the ARPES intensities and the background function that we have
employed to obtain the quasiparticle $A ({\bf k},\omega)$ for
${\bf k} = (\frac{\pi}{2},\frac{\pi}{2})$. Clearly, the peak arising from
the main valence band is very well represented, even without the
inclusion of the inelastic linear term --- similar satisfactory
agreement is found at other wave vectors. Thus, we hope that by following
this prescription we are comparing our theoretical work to the intrinsic 
spectral response of a single hole propagating in a CuO$_2$ plane in 
${\rm Sr_2CuO_2Cl_2}$.

%\newpage
\section{Theoretical Models}
\label{theory}

To study the theoretical spectrum $A({\bf k},\omega)$ of the planar copper
oxides, we start with the simple $t-J$ model with Hamiltonian,
\begin{equation}
{\cal H}_1 = -t~\sum_{nn,\sigma}
(\tilde c^\dagger_{i,\sigma}\tilde c_{j,\sigma} + {\rm H.c.})
+J~\sum_{nn}({\bf S}_i\cdot{\bf S}_j-\frac{1}{4}n_in_j)~~,
\label{H_1}
\end{equation}
where $nn$ means nearest neighbours \cite {elbioRMP}.
It is well known that this simple, one--band model cannot reproduce 
all the features of the single hole dispersion relation $E({\bf k})$
of ${\rm Sr_2CuO_2Cl_2}$ as measured by ARPES \cite{wells,lg95}.
While the $E({\bf k})$ from the $t-J$ model agrees
with experiment along the $(0,0)$ to $(\pi,\pi)$ direction,
two discrepancies remain:
(i) along the antiferromagnetic Brillouin zone (ABZ) edge
($(\pi,0)$ to $(0,\pi)$), the $t-J$ model is less dispersive; and
(ii) along the $(0,0)$ to $(\pi,0)$ direction,
the model is much more dispersive than the experimental result.

Various modifications to the $t-J$ model have been suggested to make 
the theoretical prediction in better agreement with experiment. One 
suggestion follows from the finding that a three--band model works much better
than the one--band model \cite{nazarenko,vos}, {\em viz.}
one can mimic many features of the three--band model
in the one--band model by including farther--than--nearest--neighbour
hopping terms. For example, it has become customary to include \cite {kyung}
\begin{eqnarray}
{\cal H}_2 = -t^\prime~\sum_{2nn,\sigma}&&
(\tilde c^\dagger_{i,\sigma}\tilde c_{j,\sigma} + {\rm H.c.})\nonumber\\
&&-t^{\prime\prime}~\sum_{3nn,\sigma}(\tilde c^\dagger_{i,\sigma}\tilde 
c_{j,\sigma} 
+ {\rm H.c.})~~,
\label{H_2}
\end{eqnarray}
where $2nn$ and $3nn$ are the second and third--nearest neighbours,
respectively.
Inclusion of ${\cal H}_2$ significantly lowers the anisotropy of
the effective mass tensor found in the
theoretical $E({\bf k})$ around $(\frac{\pi}{2},\frac{\pi}{2})$ \cite{kyung},
but discrepancy (ii) above still remains.
However, recently it has been suggested that so--called three--site,
spin--dependent hopping terms \cite {trugman}
will make the theoretical $E({\bf k})$ from $(0,0)$ to $(\pi,0)$
almost dispersionless \cite{sasha,wheatley}, thus eliminating this problem.
These processes are given by
\begin{eqnarray}
{\cal H}_3 = \frac{J}{4}~\sum_{j,\sigma}\sum_{\delta\ne\delta^\prime} 
(&&\tilde c^\dagger_{j+\delta,\sigma}\tilde c^\dagger_{j,-\sigma}
\tilde c_{j,\sigma}\tilde c_{j+\delta^\prime,-\sigma}\nonumber\\
&&-
\tilde c^\dagger_{j+\delta,\sigma}n_{j,-\sigma}\tilde c_{j+\delta^\prime,\sigma})~~,
\label{H_3}
\end{eqnarray}
where $\delta$ and $\delta^\prime$ are the unit vectors 
$\pm\hat{\bf x}$ and $\pm\hat{\bf y}$.

We have investigated the effects of these terms
(${\cal H}_2$ and ${\cal H}_3$) on $E({\bf k})$ of the $t-J$ model
using the same method as in Ref. \cite{lg95}. In particular,
we calculate the quasiparticle dispersion relation $E({\bf k})$
and quasiparticle weight $Z_{\bf k}$ of these models 
for a 32--site square lattice.  The parameters we employed
are $t^\prime = 0.3t$, $t^{\prime\prime} = -0.2t$, and $J=0.3t$, 
and our results are tabulated in Table \ref{zq}. 
(Note that $E({\bf k})$ as defined
in Ref.~\cite{lg95} is relative to the undoped ground
state energy.) As pointed out before \cite{lg95}, the behavior 
of $Z_{\bf k}$ of the $t-J$ model along the ABZ edge
(from $(\pi,0)$ to $(0,\pi)$) does not agree with experiment.  
ARPES data shows that the peak intensity is the largest
at $(\frac{\pi}{2},\frac{\pi}{2})$, whereas $Z_{\bf k}$  for
this wave vector for the $t-J$ is the smallest along this direction.
From Table \ref{zq} it is clear that inclusion of 
farther--than--nearest--neighbour hopping terms eliminates this
discrepancy. We also note that the three--site hopping terms ${\cal H}_3$
suppress $Z_{\bf k}$ at all {\bf k} except at the VBM
$(\frac{\pi}{2},\frac{\pi}{2})$.  Although we cannot produce a quantitative
comparison of $Z_{\bf k}$ with ARPES results, we will see
(in Section \ref{comparison}) that this suppression
of $Z_{\bf k}$ makes $A({\bf k},\omega)$ agree better with ARPES results.

\begin{figure}
\centerline{\psfig{figure=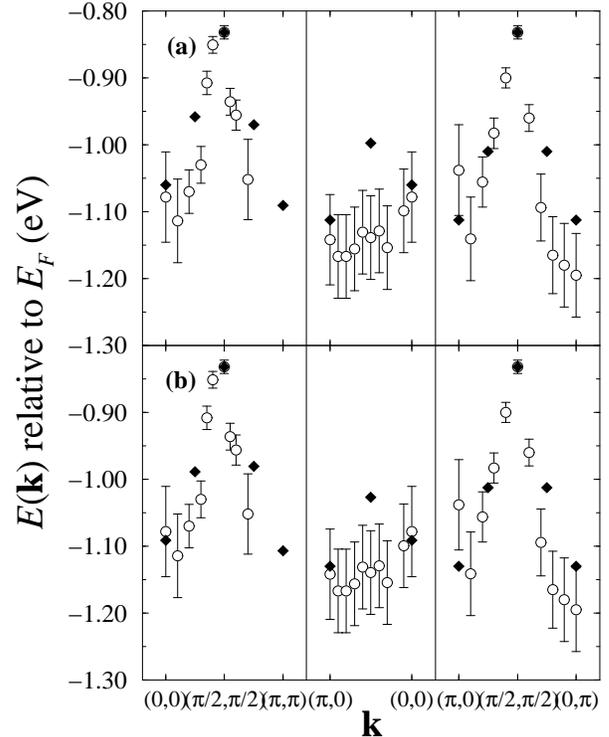,width=5.in}}
\caption{Dispersion relations for single hole propagating in
a 2D square lattice AFM described by the $t-t^\prime-t^{\prime\prime}-J$
model (a) without (${\cal H}_1 + {\cal H}_2$) and (b) with
(${\cal H}_1 + {\cal H}_2 + {\cal H}_3$) the three--site, 
spin--dependent hopping processes.  We have used the following parameters:
$t^\prime = -0.3t$, $t^{\prime\prime}=0.2t$, and $J=0.3t$. The open
circles ($\circ$) with error bars are the ARPES results from 
Ref.~\protect\onlinecite{wells}.}
\label{dispersion}
\end{figure}

To make a comparison of $E({\bf k})$ with experimental results,
we should express it in units of eV and relative to the
experimental Fermi energy.  This can be done by
using the energy scale $J=0.125$ eV and 
the location of the ARPES peak at ${\bf k}=(\frac{\pi}{2},\frac{\pi}{2})$.
In Fig.~\ref{dispersion} we plot the dispersion relation
of the ${\cal H}_1 + {\cal H}_2 + {\cal H}_3$ with and without the
three--site hopping terms, and our numerical results
are compared with ARPES data \cite{wells}. Consistent with previous 
studies \cite{kyung},
we find that adding the farther--than--nearest--neighbour hopping terms
to the $t-J$ model leads to an (essentially) isotropic effective
mass tensor near the VBM.
However, there is still some dispersion along the
$(0,0)$ to $(\pi,0)$ direction, in contrast to the featureless ARPES results.
Previous studies \cite{sasha,wheatley} have suggested that including the 
three--site
hopping term ${\cal H}_3$ can reduce the dispersion along this
direction, hence making the theoretical prediction closer
to the experimental result.
However, from Fig.~\ref{dispersion}(b) we see that ${\cal H}_3$
only lowers the energy of $(\frac{\pi}{2},0)$ by a small amount
(and thus our exact, unbiased, numerical results are in disagreement
with earlier analytical work, as well as with work on smaller
clusters).
Despite the lack of perfect agreement along this direction, this is the 
best comparison to the dispersion relation inferred from the ARPES data
for the ${\cal H}_1 + {\cal H}_2 + {\cal H}_3$ Hamiltonian that we have
found.

%\newpage
\section{Calculation of $A({\bf \lowercase{k}},\omega)$ and comparison 
with ARPES}
\label{comparison}

The principal quantity of interest in this paper is the electron spectral 
function $A({\bf k},\omega)$, and is defined by
\begin{equation}
A({\bf k},\omega) = \sum_n 
|\langle\psi_n^{N-1}|\tilde c_{{\bf k},\sigma}|\psi^N_0\rangle|^2 
\,\delta(\omega-E^N_0+E_n^{N-1})~~,
\label{Aq}
\end{equation}
where $E^N_0$ and $\psi^N_0$ are the ground--state
energy and wave function at half filling, respectively,
and $E_n^{N-1}$ and $\psi_n^{N-1}$ are the energy and wave function
of the $n$th eigenstate of the single-hole problem, respectively.
To fit the lineshape of the ARPES results
we need to broaden the delta peaks from the theoretical
calculation.  The continued fraction expansion used in Ref.~\onlinecite{lg95}
is equivalent to broadening the peaks by Lorentzians,
\begin{eqnarray}
A({\bf k},\omega) = 
\sum_n |&&\langle\psi_n^{N-1}|\tilde c_{{\bf k},\sigma}|\psi^N_0\rangle|^2 
\nonumber\\
&&\frac{1}{\pi} \biggl[ \frac{\Gamma}{(\omega-E^N_0+E_n^{N-1})^2+\Gamma^2}
\biggr]~~.
\label{lorentzian}
\end{eqnarray}
The broadening factor $\Gamma$ is a constant, independent
of {\bf k} and $\omega$.
To determine $\Gamma$
we fit $A({\bf k},\omega)$ to the ARPES at 
${\bf k}=(\frac{\pi}{2},\frac{\pi}{2})$
where the peak is most well defined.  
For the best agreement, we have to use $\Gamma=0.6t\sim 250$ meV.
Note that this is as large as the coherent bandwidth, 
which is $280\pm60$ meV.
Now, since all energy scales and $\Gamma$ are fixed
by the spectrum at ${\bf k}=(\frac{\pi}{2},\frac{\pi}{2})$, no fitting is
needed for other {\bf k}.

\begin{figure}
\centerline{\psfig{figure=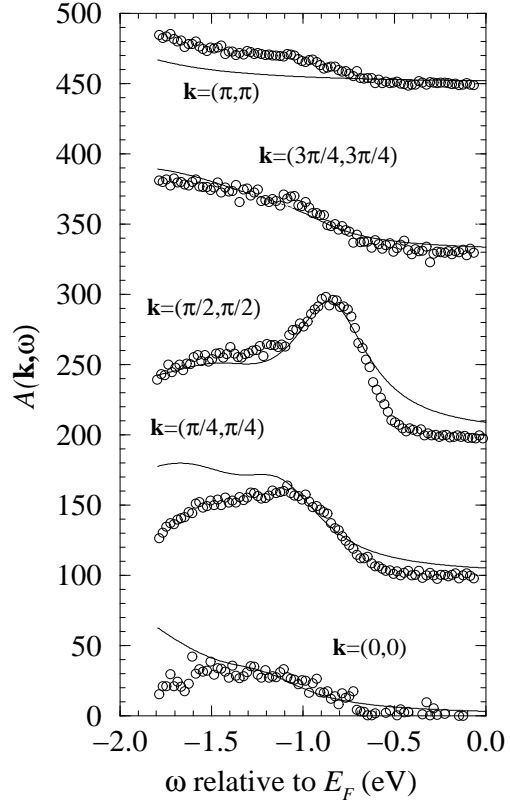,width=6.in}}
\caption{Comparison of the theoretical $A({\bf k},\omega)$ for
${\cal H}_1 + {\cal H}_2 + {\cal H}_3$ to the ARPES
data at available wavevectors {\bf k} from $(0,0)$
to $(\pi,\pi)$.  Solid lines are $A({\bf k},\omega)$ calculated
using constant broadening $\Gamma=0.6t$ 
(refer to Eq.~(\protect\ref{lorentzian}))
at the {\bf k} indicated in the graph. Open circles ($\circ$) 
are corrected ARPES data at the same {\bf k}.
(Since the experiments were not performed at exactly the
same wave vectors that we can treat numerically, for the 
theoretical ${\bf k}=(\frac{\pi}{4},\frac{\pi}{4})$
we show the ARPES data at $(0.3\pi,0.3\pi)$,
and for the theoretical ${\bf k}=(\frac{3\pi}{4},\frac{3\pi}{4})$
we show the ARPES data at $(0.8\pi,0.8\pi)$.)}
\label{constantfit}
\end{figure}

In Fig.~\ref{constantfit} we plot
$A({\bf k},\omega)$ on top of the corrected ARPES data
at the available {\bf k} from $(0,0)$ to $(\pi,\pi)$.
(These experimental data are probably the most reliable because  
in the experiment the angle between the electric field vector, the 
sample axes, and the ejected electrons remains constant.)
Since the theoretical model is supposed to describe only the low energy
physics of the CuO$_2$ plane, we truncate the spectrum at $\omega=-1.8$ eV.
It is obvious that the model can reproduce the experimental finding that 
the spectral weight is maximum at the VBM 
and that the peak broadens and weakens as {\bf k} moves away from
this point in the Brillouin zone. Further, we find that the incoherent part 
of the theoretical spectra make the peaks asymmetric, also in
qualitative agreement with experiment.  However, they are not as
asymmetric as the experimental results, and the comparison
is not entirely encouraging. Of course, the question is: is this the best 
this theoretical model can do?

Looking at the VBM peak, which is the sharpest and the most prominent, 
it is clear that we cannot completely fit the ARPES 
result with simple Lorentzians, since the high energy part of the peak falls 
off much faster than a Lorentzian. Contrary to the ARPES of metals, this faster
fall off cannot be accounted for by including a Fermi function.
To improve the comparison of our numerics and experiment, without microscopic
justification we use an energy--dependent broadening function $\Gamma(\omega)$. 
Analogous to Eq.~(\ref{lorentzian}), we write
\begin{eqnarray}
A({\bf k},\omega) = \sum_n
|&&\langle\psi_n^{N-1}|\tilde c_{{\bf k},\sigma}|\psi^N_0\rangle|^2 
\frac{1}{{\cal N}(E^N_0-E_n^{N-1})}\nonumber\\
&&\biggl[
\frac{\Gamma(\omega)}{(\omega-E^N_0+E_n^{N-1})^2+\Gamma^2(\omega)}
\biggr]~~,
\label{edependent}
\end{eqnarray}
where ${\cal N}(E)$ is the normalization factor,
\begin{equation}
{\cal N}(E)=\int^\infty_{-\infty}
\frac{\Gamma(\omega)}{(\omega-E)^2+\Gamma^2(\omega)}\,d\omega~~.
\end{equation}
From the spectrum at $(\frac{\pi}{2},\frac{\pi}{2})$ in Fig.~\ref{constantfit}, 
the low energy tail (incoherent part) of the spectrum seems to need larger 
broadening.  For simplicity we choose $\Gamma(\omega)$ to be
linear in $\omega$:
\begin{equation}
\label{edepepsilon}
\Gamma(\omega)={\rm max}(0,-0.176\ {\rm eV} - 0.55 \omega)~~.
\end{equation}
Note that at the first peak ($\omega\sim-0.85$ eV), $\Gamma(\omega)\sim290$ meV
which is similar to the constant $\Gamma$ mentioned above.

\begin{figure}
\centerline{\psfig{figure=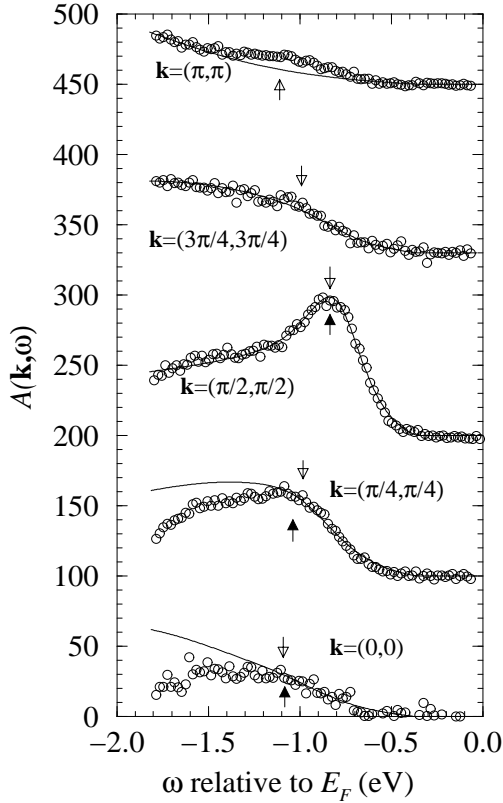,width=6.in}}
\caption{This figure shows the comparison of our spectral functions to
wave vectors along $(k,k)$, the same as in Fig.~\protect\ref{constantfit}, 
except that $A({\bf k},\omega)$ is calculated using 
$\Gamma(\omega)={\rm max}(0,-0.176\ {\rm eV} - 0.55 \omega)$. We have also
included solid arrows to indicate the positions of the first peaks of the ARPES
data as shown in Fig.~\protect\ref{dispersion},
and hollow arrows to indicate the numerically determined $E({\bf k})$.}
\label{bestfit}
\end{figure}

\begin{figure}
\centerline{\psfig{figure=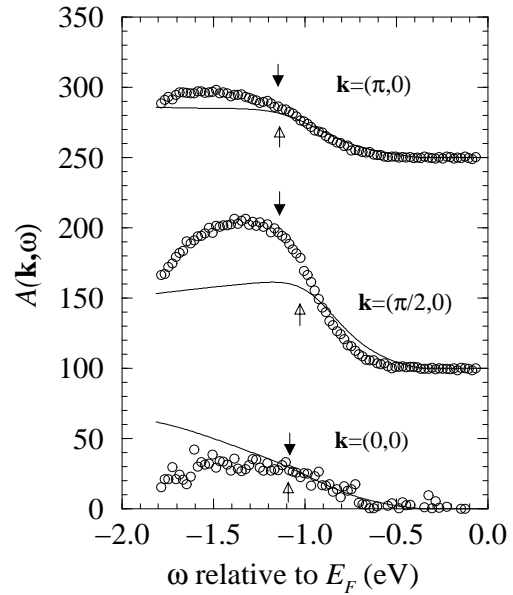,width=6.in}}
\caption{Same as Fig.~\protect\ref{bestfit} but
along the $(0,0)$ to $(\pi,0)$ direction.}
\label{00-10}
\end{figure}

\begin{figure}
\centerline{\psfig{figure=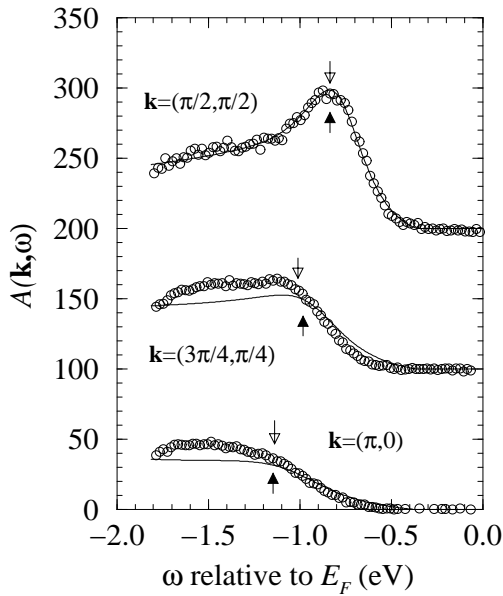,width=6.in}}
\caption{Same as Fig.~\protect\ref{bestfit} but
along the $(\pi,0)$ to $(0,\pi)$ direction. (Similar to the circumstances
discussed in Fig.~\protect\ref{constantfit}, 
for the theoretical ${\bf k}=(\frac{3\pi}{4},\frac{\pi}{4})$
we show the ARPES results for ${\bf k} = (0.3\pi,0.7\pi)$, the latter
found from an average of the data
for $(0.7\pi,0.3\pi)$ and $(0.3\pi,0.7\pi)$.)}
\label{01-10}
\end{figure}

In Fig.~\ref{bestfit} we plot
$A({\bf k},\omega)$ calculated from Eqs.~(\ref{edependent})
and (\ref{edepepsilon}) on top of the corrected ARPES data
for those {\bf k} from $(0,0)$ to $(\pi,\pi)$.
This shows that our energy--dependent broadening function
is able to produce a much improved fit at all {\bf k}.
In Figs.~\ref{00-10} and \ref{01-10}  we plot the
comparison along the $(0,0)$ to $(\pi,0)$,
and the $(\pi,0)$ to $(0,\pi)$ directions, respectively.
Figures~\ref{bestfit}, \ref{00-10}, and \ref{01-10}
show that our model fits the ARPES lineshape at many {\bf k} very well. 
The least satisfactory fit is at ${\bf k}=(\frac{\pi}{2},0)$,
which is also the least satisfactory {\bf k} point in the dispersion
relation comparison, as seen in Fig.~\ref{dispersion}.
Despite this, our model does produce the largest
spectral weight ({\em c.f.} Table~\ref{zq}) along
the $(0,0)$ to $(\pi,0)$ direction at this wave vector, similar
to experiment.

In principle, we can allow the broadening function to
be dependent on {\bf k}, $\Gamma({\bf k},\omega)$ \cite{previous_expts}.
For example, instead of determining the two coefficients in
Eq.(\ref{edepepsilon}) using the spectrum at $(\frac{\pi}{2},\frac{\pi}{2})$,
we can determine them at each {\bf k}.
We have completed such fittings, but we do not find much better results, and 
in particular, the discrepancy between theoretical and experimental
spectra at $(\frac{\pi}{2},0)$ remains.

%\newpage
\section{conclusions}
\label{concl}

To summarize, to compare the theoretical and experimental lineshapes
we have argued that the ARPES data of ${\rm Sr_2CuO_2Cl_2}$ is dominated by
direct transitions, and thus with the valence band contribution
appropriately subtracted off, one can obtain the spectral function for
a single hole propagating in a CuO$_2$ plane.
From these experimental spectral functions it is obvious
that the spectrum at each {\bf k} cannot be fit by
a single, narrow peak; instead, the incoherent part is important.

We have studied the effects of including
farther--than--nearest neighbour and three--site, spin--dependent
hopping terms to the $t-J$ model.
Consistent with previous studies, we find that these
terms increase the dispersion along the antiferromagnetic
Brillouin zone edge ($(\pi,0)$ to $(0,\pi)$), and suppress
that along the $(\pi,0)$ to $(0,0)$ direction.
In addition, these terms suppress the quasi--particle
weight except at the valence band maximum $(\frac{\pi}{2},\frac{\pi}{2})$.
This is shown to be essential in fitting the theoretical
lineshape to the experimental results.

We also find that if we use an energy--dependent broadening function
$\Gamma(\omega)$, we can obtain a reasonably good fit in
an energy range of up to 1.8 eV below the Fermi level. The fact
that we find a damping linear in energy should not be seen as direct
support for the marginal Fermi liquid hypothesis, since we are studying
an insulator and thus we cannot predict that this linear term survives
in the (anomalous) metallic state.

The least satisfactory fit is found along the $(\pi,0)$
to $(0,0)$ direction, and at $(\frac{\pi}{2},0)$ our calculated spectrum is
significantly lower than the ARPES result. To date, only three--band
models have provided a reasonable agreement with the dispersion near this point.

An interesting question that follows from our results is: What is
the origin of the incoherent processes that contribute so strongly
to spectral functions of a single hole propagating in a strongly correlated,
half--filled state? Two possibilities seem worthy
of further study. One, long--wavelength spin excitations, excluded from our
work due to our use of a finite cluster, are important. Two,
some other excitations, {\em e.g.} lattice waves, are strongly
coupled to either the hole motion or the spin waves, or possibly
both, and provide an effective energy--dependent damping of
these quasiparticles.

\acknowledgements
We wish to thank Sasha Chernyshev for a number of insightful comments.
Helpful comments from David Johnston and Bob Birgeneau are appreciated.
This work was supported by Hong Kong RGC grant HKUST619/95P (PWL) 
and the NSERC of Canada (RJG).
Numerical diagonalizations of the 32--site system were performed on the
Intel Paragon at HKUST.

% now the references. delete or change fake bibitem. delete next three
%   lines and directly read in your .bbl file if you use bibtex.


% \begin{table}
% \caption{}
% \label{}
% \end{tabular}
% \end{table}
\begin{table}
\caption{Quasiparticle weight $Z_{\bf k}$ and energy $E({\bf k})$
of the $t-t^\prime-t^{\prime\prime}-J$ model with (${\cal H}_1+{\cal H}_2+{\cal H}_3$) and
without (${\cal H}_1+{\cal H}_2$)
the three--site, spin--dependent hopping terms, calculated on a 32--site 
square lattice. The parameters that we used are $t'=-0.3t$, $t''=0.2t$, and 
$J=0.3t$.  $E({\bf k})$ is in units of $t$ and is measured relative to the
half--filled ground state energy $E_0 = -11.3297t$.}
\label{zq}
\begin{tabular}{r@{}l|cc|cc}
\multicolumn{2}{c|}{}&\multicolumn{2}{c|}{${\cal H}_1+{\cal H}_2$}&
\multicolumn{2}{c}{${\cal H}_1+{\cal H}_2+{\cal H}_3$}\\
%\cline{3-6}
\multicolumn{2}{c|}{\bf k}&$Z_{\bf k}$&$E({\bf k})$&$Z_{\bf k}$&$E({\bf k})$\\
\tableline
(0,&0)                                  & 0.0916 & 1.8390 & 0.0195 & 2.0668\\
($\frac{\pi}{4}$,&$\frac{\pi}{4}$)      & 0.1820 & 2.0832 & 0.0462 & 2.3120\\
($\frac{\pi}{2}$,&$\frac{\pi}{2}$)      & 0.3527 & 2.3868 & 0.3721 & 2.6880\\
($\frac{3\pi}{4}$,&$\frac{3\pi}{4}$)    & 0.0749 & 2.0554 & 0.0379 & 2.3314\\
($\pi$,&$\pi$)                          & 0.0021 & 1.7661 & 0.0000 & 2.0280\\
($\pi$,&$\frac{\pi}{2}$)                & 0.1086 & 1.9741 & 0.0736 & 2.2418\\
($\pi$,&0)                              & 0.0287 & 1.7126 & 0.0276 & 1.9733\\
($\frac{\pi}{2}$,&0)                    & 0.2117 & 1.9897 & 0.1150 & 2.2202\\
($\frac{3\pi}{4}$,&$\frac{\pi}{4}$)     & 0.0721 & 1.9588 & 0.0340 & 2.2556\\
\end{tabular}
\end{table}

\end{document}